\documentclass[11pt]{article}
\usepackage{epsfig,makeidx,amsmath,amsfonts,latexsym,bm}
\newcommand{\E}{{\mathbb E}}
\newcommand {\PP}{\mathbb P}

\begin{document}

\title{Growing networks with preferential deletion and\newline addition of edges}

\author{Maria Deijfen \thanks{Department of Mathematics, Stockholm University, 106 91 Stockholm. Email: mia@math.su.se. Work carried out during a visit to the Institut Mittag-Leffler.}\and Mathias Lindholm \thanks{Department of Mathematics, Stockholm University, 106 91 Stockholm. Email: lindholm@math.su.se. Work carried out during a visit to the Institut Mittag-Leffler.}}

\date{May 2009}

\maketitle

\thispagestyle{empty}

\begin{abstract}

\noindent A preferential attachment model for a growing network incorporating deletion of edges is studied and the expected asymptotic degree distribution is analyzed. At each time step $t=1,2,\ldots$, with probability $\pi_1>0$ a new vertex with one edge attached to it is added to the network and the edge is connected to an existing vertex chosen proportionally to its degree, with probability $\pi_2$ a vertex is chosen proportionally to its degree and an edge is added between this vertex and a randomly chosen other vertex, and with probability $\pi_3=1-\pi_1-\pi_2<1/2$ a vertex is chosen proportionally to its degree and a random edge of this vertex is deleted. The model is intended to capture a situation where high-degree vertices are more dynamic than low-degree vertices in the sense that their connections tend to be changing. A recursion formula is derived for the expected asymptotic fraction $p_k$ of vertices with degree $k$, and solving this recursion reveals that, for $\pi_3<1/3$, we have $p_k\sim k^{-(3-7\pi_3)/(1-3\pi_3)}$, while, for $\pi_3>1/3$, the fraction $p_k$ decays exponentially at rate $(\pi_1+\pi_2)/2\pi_3$. There is hence a non-trivial upper bound for how much deletion the network can incorporate without loosing the power-law behavior of the degree distribution. The analytical results are supported by simulations.

\vspace{0.5cm}

\noindent \emph{Keywords:} Preferential attachment, preferential deletion, complex networks, random graphs, degree distribution.

\vspace{0.5cm}

\noindent PACS 2008 classification: 89.75.Hc; 89.75.Da

\end{abstract}

\section{Introduction}

During the last decade there have been tremendous advances in the study of large complex networks; see \cite{AlB-survey,DorMen03,NewMan-structfunc} for surveys. Empirical observations on real networks such as social contact structures, citation networks and the Internet/World Wide Web have revealed that many networks share similar properties, in particular, they typically have power-law degree sequences, that is, the number of vertices with degree $k$ decays as an inverse power of $k$; see e.g. \cite{FFF99,Liljeros,Redner}. One way of modeling networks is by using random graphs, and large efforts have been made to formulate graph models that can capture the properties of real networks. The classical Erd{\H o}s-R{\' e}nyi graph, see \cite{ER1}, is discarded in this context because of its much too simplified structure. A specific point where it fails to capture the empirical observations is the degree distribution (which is a Poisson distribution in the limit in the Erd{\H o}s-R{\' e}nyi graph). To obtain a richer structure, the Erd{\H o}s-R{\' e}nyi model has been generalized in various ways, for instance by making the edge probabilities inhomogeneous, see e.g.\ \cite{BJR,BDM-L05}. Other model types that have been studied include the configuration model, see \cite{MR1,MR2}, and certain bipartite graphs, see \cite{DK,NewMan-cluster}.

The models mentioned above are all static in the sense that the number of vertices is taken to be constant (albeit large). In reality however many networks are dynamic: the number of vertices increases as new vertices arrive in the network and decays as vertices leave the network. A framework for growing networks, based on the idea that new vertices tend to connect to high-degree vertices rather than low-degree vertices, is provided by preferential attachment models, introduced in the context of complex network modeling by Barabasi and Albert \cite{BarA-science} in 1999 to describe the World Wide Web. Barabasi and Albert observed that such a mechanism leads to power-law degree sequences, and their findings have later been confirmed and elaborated on in \cite{BRST01}. The model type has received a lot of attention, see e.g.\ \cite{BR03,Durr_rgd,Mitz} for surveys, and the growth mechanism has been generalized in various ways, see e.g.\ \cite{cf} for a very general version of the model.

There is considerably less work on preferential attachment based models where edges and/or vertices also can be removed. In \cite{DorMen01} a model is studied where edges between existing vertices are added or removed each time a vertex is added. Addition and deletion can however not be incorporated simultaneously in this model. Later, Cooper et al.\ \cite{CFV} and Chung and Lu \cite{CL-del} have studied models where vertices and edges are deleted randomly, and shown that this leads to power-law degree sequences with exponents that depend on the addition/removal probabilities. Furthermore, Moore et al.\ \cite{Moore} and Cami and Deo \cite{DeoC} analyze models where vertices are deleted.

In this paper we consider a model where edges can be deleted and new edges created between existing vertices, the deletion and addition of edges occurring at vertices chosen preferentially proportionally to their degree. The edge deletion/addition is hence done preferentially with a vertex focus: we pick the vertex to lose/gain an edge. This is intended to describe a scenario where high-degree vertices tend to have more dynamic connections as compared to low-degree vertices, that is, the connections of high-degree vertices are more prone to break, but high-degree vertices are also more likely to create new connections to existing vertices. It is not hard to come up with real networks where this is likely to be the case. In a social network for instance, very social individuals with a large number of contacts may have weaker bonds to their acquaintances as compared to  individuals with very few contacts. On the other hand, the fact that social individuals have many contacts makes it more likely for them to come in contact with other people in the population and create new bonds as well.

In order for the model to define a growing network, the edge removal probability, denoted by $\pi_3$, is assumed to be smaller than 1/2. Interestingly, it turns out that the model exhibits a transition at $\pi_3=1/3$: for $\pi_3<1/3$ the expected degree distribution obeys a power-law with an exponent that diverges as $\pi_3$ approaches 1/3, while, for $\pi_3>1/3$, the degree distribution decays exponentially. There is hence a non-trivial limit for the removal probability above which the power-law behavior of the degrees is lost. Furthermore, the exponent in the power-law regime of the model turns out to depend on the probability $\pi_1$ of adding a vertex and the probability $\pi_2$ of adding an edge between existing vertices only through their sum, that is, the relation between the vertex and the edge addition probabilities does not affect the tail behavior of the degree distribution (although it does affect the mean degree of course). 

\section{The model}

The model that we will consider is described by a graph process $\{G(t)\}_{t\geq 1}$ consisting of simple graphs, that is, graphs without self-loops and multiple edges. Write $V_t$ and $E_t$ for the set of vertices and edges, respectively, in $G(t)$ and let $v_t=|V_t|$ and $e_t=|E_t|$. The graph $G(1)$ is taken to be a graph consisting of two vertices connected by an edge and, for $t\geq 1$, the graph $G(t+1)$ is constructed from $G(t)$ in that one of the following three steps is carried out:
\begin{itemize}
\item[1.] With probability $\pi_1>0$, a vertex with one edge attached to it is added to the graph. The edge is connected to an existing vertex $w$ with probability proportional to degree, that is,
    $$
    \PP(w=u)=\frac{d_t(u)}{2e_t}\quad\textrm{for }u\in V_t,
    $$
    where $d_t(u)$ denotes the degree of vertex $u$ in $G(t)$.
\item[2.] With probability $\pi_2$, an edge is added between a vertex chosen proportionally to its degree and another vertex chosen uniformly at random. If this results in a multiple edge between two vertices, the edges are merged.
\item[3.] With probability $\pi_3=1-\pi_1-\pi_2$, an edge is deleted from the graph in that a vertex is chosen proportionally to its degree and an edge chosen uniformly at random from the set of all edges incident to this vertex is removed.
\end{itemize}

\noindent When $e_t=0$, modifications of the rules are needed: for a step of type 1, the edge is connected to a randomly chosen vertex, for a step of type 2, an edge is added between two randomly chosen vertices and, for a step of type 3, the graph is left unchanged. However, throughout, we will assume that $\pi_3<1/2$ so that the number of edges in the graph is indeed growing (on average). This means that a graph without any edges occurs very rarely and how we deal with this case should not matter for the results derived here. Indeed, the graph $G(t)$ can have $e_t=0$ only if at least $\lceil t/2\rceil$ steps have been of type 3 and, for $\pi_3<1/2$, the probability of this decays exponentially in $t$. We can hence disregard this possibility in the below derivations.

Note that, when $\pi_2=0$, a vertex that ends up having degree 0 at some point will, with a probability tending to 1, keep on having degree 0 forever. When $\pi_2>0$ on the other hand, a vertex with degree 0 can acquire new edges (by being selected as an endpoint of an edge added as a result of step 2). Also note that the step 3 is equivalent to deleting an edge uniformly at random from the set of edges: First a vertex $u$ is chosen with probability $d_t(u)/2e_t$ and then an edge incident with u is chosen randomly, that is, each incident edge has probability $1/d_t(u)$ of being chosen. The probability that a given edge in the graph is chosen via any of its two endpoints is hence $1/e_t$.

\section{Expected asymptotic degree distribution}

Write $N_k(t)$ for the number of vertices with degree $k$ in $G(t)$ and let $\E[\cdot]$ denote mathematical expectation. Assume that $\E[N_k(t)]/t$ converges to some limit $p_k$ as $t\to\infty$. In view of previous work on related models, e.g.\ \cite{cf} and \cite{CFV}, this assumption is indeed reasonable. It can presumably be confirmed rigorously by a more careful analysis of (\ref{bet_vv}) along similar lines as in \cite{CFV}\footnote{We have that $\E[N_k(t)]/t$ satisfies the recursion obtained from (\ref{bet_vv}) conditional on that the number of vertices and edges respectively in the graph are concentrated around their means. With sufficient concentration bounds at hand it would then follow that the solution of the recursion, which we denote by $p_k$, is a good approximation of $\E[N_k(t)]/t$ for large $t$ so that, in particular, $\E[N_k(t)]/t\to p_k$; c.f.\ \cite[Lemma 5.2]{CFV}.}. We will analyze the behavior of $p_k$ as $k\to\infty$.

\subsection{A recursion formula for $p_k$}

Given $G(t)$, the average change in the $(t+1)$:th step in the number of vertices of degree $k$ ($k=0,1,\ldots$) is given by
\begin{equation}\label{bet_vv}
\E[N_k(t+1)-N_k(t)|G(t)]=\pi_1C_{k,t}^{(1)}+\pi_2C_{k,t}^{(2)}+\pi_3C_{k,t}^{(3)}+
\pi_1\mathbf{1}_{\{k=1\}},
\end{equation}

\noindent where $C_{k,t}^{(i)}$ denotes the average change if the step is of type $i$ ($i=1,2,3$). Defining $N_{-1}(t):=0$ for all $t$, we have
\begin{eqnarray*}
C_{k,t}^{(1)} & = & \frac{(k-1)N_{k-1}(t)}{2e_t}-\frac{kN_k(t)}{2e_t}\\
C_{k,t}^{(2)} & = & \frac{(k-1)N_{k-1}(t)}{2e_t}+ \frac{N_{k-1}(t)}{v_t}-\frac{kN_k(t)}{2e_t}-\frac{N_k(t)}{v_t}\\
C_{k,t}^{(3)} & = & \frac{2(k+1)N_{k+1}(t)}{2e_t}-\frac{2kN_k(t)}{2e_t}.
\end{eqnarray*}

\noindent Here, in deriving the expression for $C_{k,t}^{(3)}$, we have used the fact that the fraction of neighbors with degree $k$ of the vertex chosen in step 3 is given by
$$
\sum_{u\in V_t}\frac{d_t(u)}{2e_t}\cdot\frac{N_k(t;u)}{d_t(u)}=\frac{1}{2e_t}\sum_{u\in V_t}N_k(t;u)=\frac{kN_k(t)}{2e_t},
$$
where $N_k(t;u)$ is the number of neighbors with degree $k$ of the vertex $u$ at time $t$ (also recall that $d_t(u)$ denotes the degree of $u$ at time $t$). To obtain a recursion for $p_k$, we take expectation in (\ref{bet_vv}) and let $t\to\infty$. Ignoring the effects of merging multiple edges in step 2 and the modifications of the growth rule when $e_t=0$ --- indeed, both these phenomena are negligible in the limit of large $t$ --- we have that $e_t$ is a sum of $t$ independent identically distributed variables with mean $\pi_1+\pi_2-\pi_3$. This means that $\E[e_t]/t\to \pi_1+\pi_2-\pi_3$ and $e_t/t\to \pi_1+\pi_2-\pi_3$ in probability (that is, $\PP(|\frac{e_t}{t}-(\pi_1+\pi_2-\pi_3)|>\varepsilon)\to 0$ for any $\varepsilon>0$). Furthermore, we have that $N_k(t)/2e_t\leq 1/k$. Hence, assuming that the convergence of $N_k(t)/t$ to $p_k$ holds also in probability, it follows from bounded convergence (see e.g.\ \cite{Durr_pte}) that

\begin{equation}\label{kvot_vv}
\E\left[\frac{N_k(t)}{2e_t}\right]\to \frac{p_k}{2(\pi_1+\pi_2-\pi_3)}.
\end{equation}

\noindent The other terms of the same form are dealt with analogously. Similarly $\E[v_t]/t\to \pi_1$ and $v_t/t\to \pi_1$ in probability, and hence $\E[N_k(t)/v_t]\to p_k/\pi_1$. Also note that, since $\E[N_k(t)]\approx tp_k$ for large $t$, we have $\E[N_k(t+1)-N_k(t)]\approx (t+1)p_k-tp_k=p_k$. Merging terms together and defining $p_{-1}:=0$, this leads to the recursion
\begin{equation}\label{rec}
[\alpha_2(k+2)+\beta_2]p_{k+2}+[\alpha_1(k+1)+\beta_1]p_{k+1}+[\alpha_0k+\beta_0]p_k=
\pi_1^2(\pi_1+\pi_2-\pi_3)\mathbf{1}_{\{k=0\}}\quad\textrm{for }k\geq -1,
\end{equation}
where
$$
\begin{array}{ll}
\alpha_2=2\pi_1\pi_3 & \beta_2=0\\
\alpha_1=-\pi_1(\pi_1+\pi_2+2\pi_3) & \beta_1=-2(\pi_1+\pi_2)(\pi_1+\pi_2-\pi_3)\\
\alpha_0=\pi_1(\pi_1+\pi_2) & \beta_0=2\pi_2(\pi_1+\pi_2-\pi_3).
\end{array}
$$

To get a first idea of the asymptotic behavior of the solution to (\ref{rec}),
divide both sides by $(k+1)p_{k+1}$ and take the limit as $k\to \infty$. Writing $\lim_{k\to\infty}p_{k+1}/p_k=x$, this yields the equation $\alpha_2x^2+\alpha_1x+\alpha_0=0$, which is solved by $x=1$ and $x=\alpha_0/\alpha_2=(\pi_1+\pi_2)/2\pi_3$. Here $x=1$ corresponds to a power-law tail, while, for $\pi_1+\pi_2<2\pi_3$, the latter solution corresponds to exponential decay at rate $(\pi_1+\pi_2)/2\pi_3$. We now analyze the solution of (\ref{rec}) in more detail.

\subsection{Solving the recursion}

The recursion formula (\ref{rec}) is of second order and hence, imposing the two conditions $\sum p_k=1$ and $p_{-1}=0$ yields a unique solution. We will use the method of Laplace, see e.g.\ \cite[p.579]{Jordan}, to solve the homogeneous equation
\begin{equation}\label{hrec}
[\alpha_2(k+2)+\beta_2]p_{k+2}+[\alpha_1(k+1)+\beta_1]p_{k+1}+[\alpha_0k+\beta_0]p_k=0\quad\textrm{for }k\geq 1.
\end{equation}
The solution space of this equation is two-dimensional but typically only one of the solutions is convergent. It can then be seen along the same lines as in \cite[Lemma 6.2]{CFV} that, if $p_k$ is a convergent solution of (\ref{hrec}), then there is a constant $c>0$ such that $cp_k$ solves the full recursion (\ref{rec}) under the conditions $\sum p_k=1$ and $p_{-1}=0$. To analyze the tail behavior of the solution we can hence restrict attention to the solution of the homogeneous equation.

Following the method of Laplace, we substitute
\begin{equation}\label{Lap_sub}
p_k=\int_a^bt^{k-1}h(t)dt,
\end{equation}
and note that integration by parts of \eqref{Lap_sub} yields
\begin{equation}\label{Lap_part_int}
kp_k=[t^kh(t)]_a^b-\int_a^bt^kh'(t)dt.
\end{equation}
Furthermore, define
$$
\begin{array}{l}
\psi_\alpha(t):=\alpha_2t^2+\alpha_1t+\alpha_0=\alpha_2(1-t)\left(\frac{\alpha_0}{\alpha_2}-t\right)\\
\psi_\beta(t):=\beta_2t^2+\beta_1t+\beta_0=\beta_1t+\beta_0.
\end{array}
$$
By substituting \eqref{Lap_sub} and \eqref{Lap_part_int} into \eqref{hrec} it can be seen that the homogeneous equation (\ref{hrec}) is satisfied if $a,b$ and $h(t)$ are chosen such that
\begin{equation}\label{cond1}
[t^kh(t)\psi_\alpha(t)]_a^b=0
\end{equation}
and
\begin{equation}\label{cond2}
\frac{h'(t)}{h(t)}=\frac{\psi_\beta(t)}{t\psi_\alpha(t)}.
\end{equation}
The differential equation (\ref{cond2}) can be integrated and is solved by
$$
h(t)=t^{\gamma_1}(1-t)^{\gamma_2}\left(\frac{\alpha_0}{\alpha_2}-t\right)^{\gamma_3}
$$
where
$$
\begin{array}{l}
\gamma_1=\frac{\beta_0}{\alpha_0}=\frac{2\pi_2(\pi_1+\pi_2-\pi_3)}{\pi_1(\pi_1+\pi_2)}\\
\gamma_2=\frac{\beta_0+\beta_1}{\alpha_2-\alpha_0}=
\frac{2(\pi_1+\pi_2-\pi_3)}{\pi_1+\pi_2-2\pi_3}\\
\gamma_3=-\frac{\beta_0\alpha_2+\beta_1\alpha_0}{\alpha_0(\alpha_2-\alpha_0)}=
\frac{2(\pi_1+\pi_2-\pi_3)(2\pi_2\pi_3-(\pi_1+\pi_2)^2)}{\pi_1(\pi_1+\pi_2)(\pi_1+\pi_2-2\pi_3)}.
\end{array}
$$
\noindent Note that $\gamma_1+\gamma_2+\gamma_3=0$ and $\gamma_1\geq 0$.\medskip

\noindent \textbf{The power-law regime.} Assume that $\pi_3<1/3$, that is, $2\pi_3<\pi_1+\pi_2$, so that equivalently $\alpha_0/\alpha_2>1$. Then $\gamma_2>0$ and (\ref{cond1}) is satisfied with $a=0$ and $b=1$. Using the notation $\asymp$ to denote that the left hand side is bounded from above and below by constants times the right hand side, and the notation $\sim$ to denote that the quotient of the right hand side and the left hand side converges to a constant, we get that
\begin{eqnarray*}
p_k & = & \int_0^1t^{k-1}t^{\gamma_1}(1-t)^{\gamma_2}\left(\frac{\alpha_0}{\alpha_2} -t\right)^{\gamma_3}dt\\
 & \asymp & \int_0^1t^{k+\gamma_1-1}(1-t)^{\gamma_2}dt\\
 & = & \frac{\Gamma(k+\gamma_1)\Gamma(1+\gamma_2)}{\Gamma(k+1+\gamma_1+\gamma_2)}\\
 & \sim & k^{-(1+\gamma_2)}.
\end{eqnarray*}

\begin{figure}[h!]
\begin{center}
\mbox{{\epsfig{file=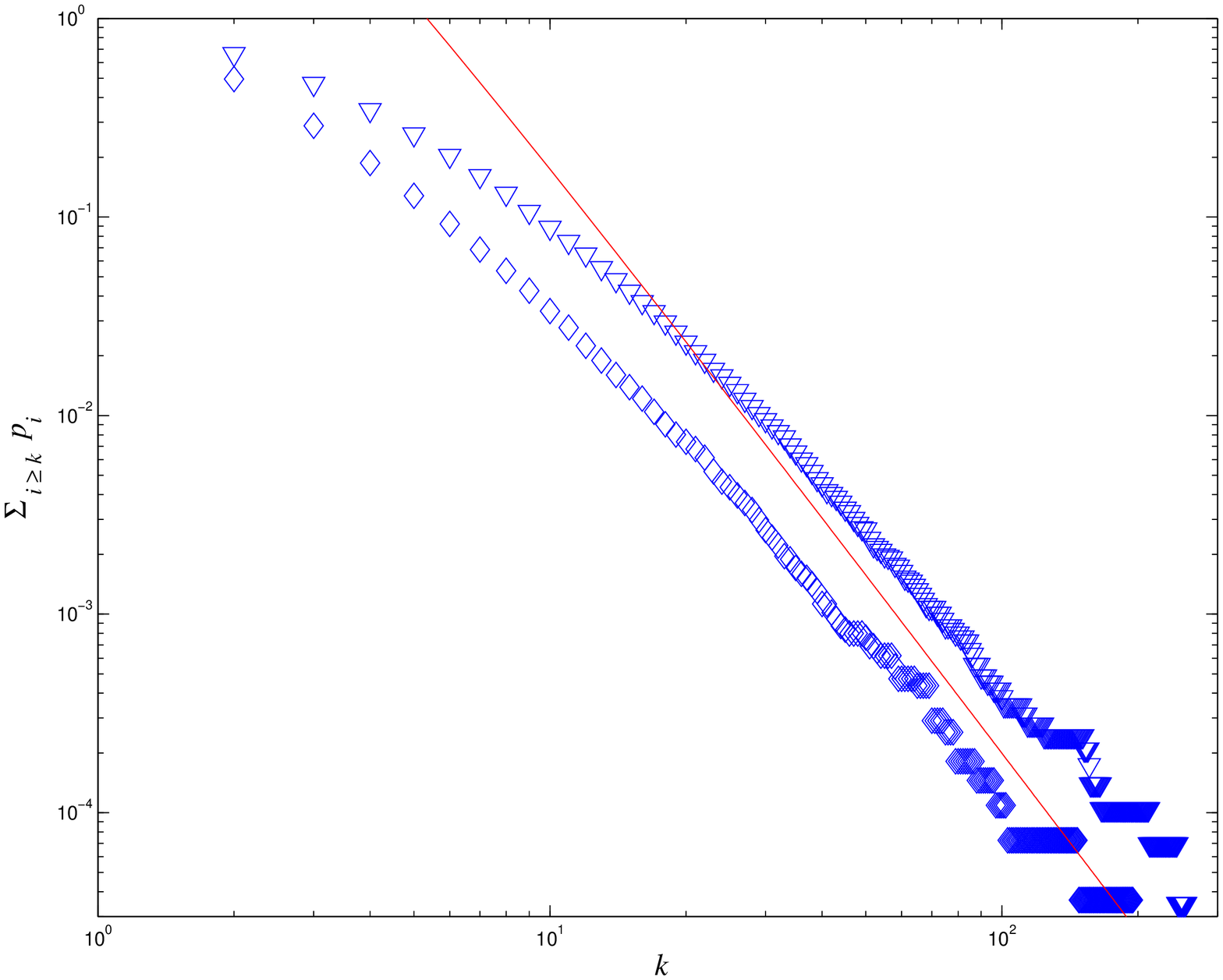,width=0.7\textwidth,height=0.5\textheight}}}
\end{center}
\caption{Simulation of the model in the power-law regime. The figure shows the tail probabilities of the degree distribution when $\pi_1=0.3,\pi_2=0.5$ and $\pi_3=0.2$ (triangles), and $\pi_1=0.5,\pi_2=0.3$ and $\pi_3=0.2$ (diamonds) together with the theoretical power-law tail (solid line). These choices of $\pi$ both give power-law exponent 4. Each of the simulated curves correspond to a single realization of the process with the number of time steps taken such that the expected number of nodes equals 30,000.}\label{tung_svans}
\end{figure}

\noindent Here
$$
1+\gamma_2=\frac{3-7\pi_3}{1-3\pi_3},
$$
which is monotonically increasing on the interval $\pi_3\in[0,1/3)$. For $\pi_3=0$ we get $1+\gamma_2=3$, and the exponent then diverges as $\pi_3\nearrow 1/3$. Figure \ref{tung_svans} shows simulations of the model for two different sets of parameter values, both with $\pi_3<1/3$. As can be seen, the agreement with the analytical prediction is good. The reason for plotting the tail probabilities of the degree distribution is to reduce the noise. \medskip

\begin{figure}[h!]
\begin{center}
\mbox{{\epsfig{file=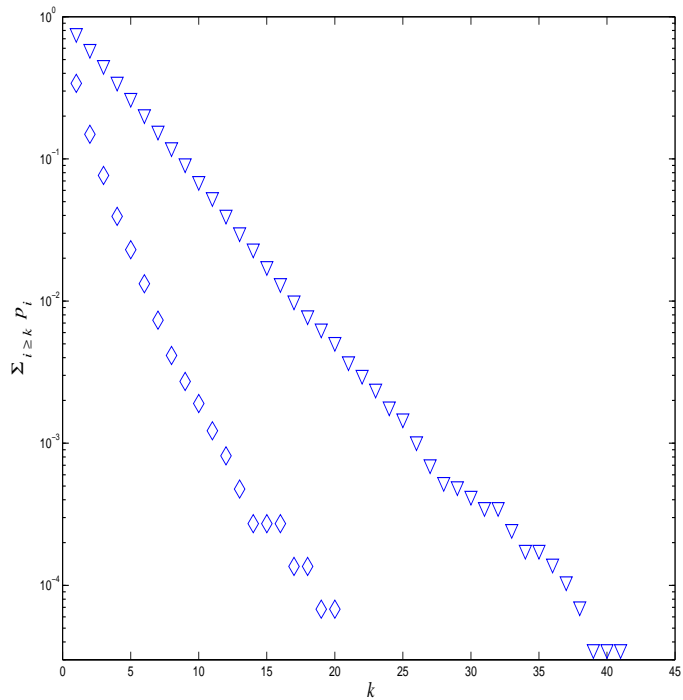,width=0.7\textwidth,height=0.5\textheight}}}
\end{center}
\caption{Simulation of the model in the exponential decay region. The figure shows the tail probabilities of the degree distribution when $\pi_1=0.1,\pi_2=0.5$ and $\pi_3=0.4$ (triangles), and $\pi_1=0.5,\pi_2=0.1$ and $\pi_3=0.4$ (diamonds). Each of the simulated curves correspond to a single realization of the process with the number of time steps taken such that the expected number of nodes equals 30,000.}\label{exp_svans}
\end{figure}

\noindent \textbf{Exponential decay.} Now assume that $\pi_3>1/3$, that is, $2\pi_3>\pi_1+\pi_2$ so that hence $\alpha_0/\alpha_2<1$ and $\gamma_2<0$. Furthermore, assume that $\gamma_3>-1$. Then (\ref{cond1}) is satisfied with $a=0$ and $b=\alpha_0/\alpha_2$ and we get
\begin{eqnarray*}
p_k & = & \int_0^{\alpha_0/\alpha_2}t^{k+\gamma_1-1}(1-t)^{\gamma_2} \left(\frac{\alpha_0}{\alpha_2}-t\right)^{\gamma_3}dt\\
& \asymp & \left(\frac{\alpha_0}{\alpha_2}\right)^{k+\gamma_3}\int_0^1y^{k+\gamma_1-1}(1-y)^{\gamma_3}dy\\
& \sim & \left(\frac{\alpha_0}{\alpha_2}\right)^k k^{-(1+\gamma_3)},
\end{eqnarray*}

\noindent that is, $p_k$ decays exponentially at rate $\alpha_0/\alpha_2=(\pi_1+\pi_2)/2\pi_3$, with an extra power-law factor $k^{-(1+\gamma_3)}$. If the extra assumption that $\gamma_3>-1$ is not satisfied the above described method will instead result in that $p_k$ can be expressed in terms of a certain contour integral. A detailed description of the method in the contour integral case together with the corresponding analysis is given in \cite[Section 4]{cf}. By using \cite[Eq.\ (24)]{cf} we can again ascertain that $p_k\sim (\alpha_0/\alpha_2)^k\,k^{-(1+\gamma_3)}$ for large values of $k$. Figure \ref{exp_svans} shows simulations of the model for two different parameter values with $\pi_3>1/3$.



\subsection{Special cases.}

Let $\pi_2=0$ and write $\pi_1=p$ and $\pi_3=1-\pi_1=q$, with $p>1/2$, that is, at each step either a vertex with one edge attached to it is added to the graph or an edge is removed. Then, for $q<1/3$, we have $p_k\sim k^{-(3-7q)/(1-3q)}$ while, for $q>1/3$, we have $p_k\sim (p/2q)^k$ (ignoring the power-law factor). This model is identical to a special case of a model treated in \cite{CFV} when setting $\alpha_0=q, \alpha_1=p$ and $\alpha=p$ in their notation. The paper \cite{CFV} is only concerned with the power-law regime, but in that situation the results agree. Another special case is when $\pi_3=0$ so that, at each time step, either a vertex with one edge attached to it is added to the graph or an edge is added. In this situation $p_k\sim k^{-3}$ regardless of the values of $\pi_1$ and $\pi_2$. This model is a sub-model of the general model treated in \cite{cf} when the parameters of the model there are set according to $\alpha=\pi_2, p_1=q_1=1, \beta=0, \delta=0$ and $\gamma=1$. Again our results agree with the existing ones. When $\pi_2=\pi_3=0$ we obtain the standard preferential attachment model with $p_k\sim k^{-3}$.

\section{Summary and discussion}

We have analyzed a dynamically evolving graph where, in addition to the standard preferential attachment step, edges are also sent out and deleted from existing vertices that are chosen proportionally to their degree. High-degree vertices hence tend to experience a higher versatility in their connections as compared to low-degree vertices. Assuming that the expected number of vertices with degree $k$ converges to a limit $p_k$ as time goes to infinity, we have analyzed the behavior of $p_k$ as $k\to\infty$. It turns out that $p_k$ decays exponentially if the removal probability exceeds 1/3, while it has a power-law tail for smaller values of the removal probability, the exponent going to infinity as the removal probability approaches 1/3. Since real networks typically exhibit power-law degree sequences (with fairly small exponents) this indicates that these networks are the outcomes of edge addition/deletion processes where the addition rate well exceeds the deletion rate.

As for future work, the results in the paper may be possible to make rigorous along similar lines as in \cite{CFV}. Concentration results seem to be difficult to establish for preferential attachment models incorporating deletion, but some recently observed connections between preferential attachment models and branching processes in continuous time might be useful in this context. It would indeed also be interesting to study other versions of preferential attachment models that incorporate deletion of edges. One option would be to assign a random life-time to each edge that is added to the network and remove the edge when this time expires, generalizing the work in \cite{Val}. One could also think of a rule where the probability of removing an edge is large just after the edge has been added, but then gradually decreases. This is to model a situation where the connections tend to stabilize --- in a social network for instance, not all new acquaintances turn into long-term friends, but a relation that has lasted for a while may be more stable.

Another line of investigation would be to look at dynamical features of the graph, such as for instance the evolution of the degree of a fixed vertex. As previously mentioned, the possibility for a vertex with degree zero of acquiring new edges depends on whether $\pi_2$ is zero or positive. It has been pointed out to us by an anonymous referee that this could lead to interesting differences in the behavior of the asymptotic degree of a fixed vertex in the exponential regime. This would be worthwhile to investigate further.

\end{document}